\documentclass[aps,prd,preprint,floatfix,nofootinbib,superscriptaddress]{revtex4}
\usepackage{graphicx,subfigure,epsfig,epsf}
\newcommand{\beq}{\begin{equation}}
\newcommand{\eeq}{\end{equation}}
\newcommand{\bea}{\begin{eqnarray}}
\newcommand{\eea}{\end{eqnarray}}
\newcommand{\bec}{\begin{center}}
\newcommand{\eec}{\end{center}}
\newcommand{\lvm}{\leavevmode}

\def\etal{{\it et al.}}

\begin{document}

\def\issue(#1,#2,#3){{\bf #1,} #2 (#3)} 
\def\APP(#1,#2,#3){Acta Phys.\ Pol.\ B \ \issue(#1,#2,#3)}
\def\ARNPS(#1,#2,#3){Ann.\ Rev.\ Nucl.\ Part.\ Sci.\ \issue(#1,#2,#3)}
\def\CPC(#1,#2,#3){Comp.\ Phys.\ Comm.\ \issue(#1,#2,#3)}
\def\CIP(#1,#2,#3){Comput.\ Phys.\ \issue(#1,#2,#3)}
\def\EPJC(#1,#2,#3){Eur.\ Phys.\ J.\ C\ \issue(#1,#2,#3)}
\def\EPJD(#1,#2,#3){Eur.\ Phys.\ J. Direct\ C\ \issue(#1,#2,#3)}
\def\IEEETNS(#1,#2,#3){IEEE Trans.\ Nucl.\ Sci.\ \issue(#1,#2,#3)}
\def\IJMP(#1,#2,#3){Int.\ J.\ Mod.\ Phys. \issue(#1,#2,#3)}
\def\JHEP(#1,#2,#3){J.\ High Energy Physics \issue(#1,#2,#3)}
\def\JPG(#1,#2,#3){J.\ Phys.\ G \issue(#1,#2,#3)}
\def\MPL(#1,#2,#3){Mod.\ Phys.\ Lett.\ \issue(#1,#2,#3)}
\def\NP(#1,#2,#3){Nucl.\ Phys.\ \issue(#1,#2,#3)}
\def\NIM(#1,#2,#3){Nucl.\ Instrum.\ Meth.\ \issue(#1,#2,#3)}
\def\PL(#1,#2,#3){Phys.\ Lett.\ \issue(#1,#2,#3)}
\def\PRD(#1,#2,#3){Phys.\ Rev.\ D \issue(#1,#2,#3)}
\def\PRL(#1,#2,#3){Phys.\ Rev.\ Lett.\ \issue(#1,#2,#3)}
\def\PTP(#1,#2,#3){Progs.\ Theo.\ Phys. \ \issue(#1,#2,#3)}
\def\RMP(#1,#2,#3){Rev.\ Mod.\ Phys.\ \issue(#1,#2,#3)}
\def\SJNP(#1,#2,#3){Sov.\ J. Nucl.\ Phys.\ \issue(#1,#2,#3)}
\bibliographystyle{revtex}

\title{Anomalous Triple Gauge Boson Couplings in $e^{-}e^{+} \rightarrow \gamma \gamma$
for Non Commutative Standard Model}

\author{N. G. Deshpande}
\email{desh@uoregon.edu}
\affiliation{Institute of Theoretical Science, University of Oregon, Eugene, Oregon 97403, USA}

\author{Sumit K. Garg}
\email{sumit@cts.iisc.ernet.in}
\affiliation{Centre for High Energy Physics, Indian Institute of Science, Bangalore 560 012, India}
%

%

\begin{abstract}
We investigate $e^{+}e^{-}\rightarrow \gamma\gamma$ 
process  within the Seiberg-Witten 
expanded noncommutative standard model(NCSM) scenario in the presence of  anomalous triple gauge
boson couplings. This study is done with and without initial
beam polarization and we restrict
ourselves to leading order effects of non commutativity i.e. $O(\Theta)$.
The non commutative(NC) corrections are sensitive to  the electric component($\vec{\Theta}_E$) of NC parameter. We include 
the effects of earth rotation in our analysis. This study is done by investigating the effects of non 
commutativity on different time averaged cross section observables. We have also defined forward backward 
asymmetries which will be exclusively sensitive to anomalous couplings. We have looked into the sensitivity 
of these couplings  at  future experiments at the International Linear Collider(ILC). This analysis is done 
under realistic ILC conditions with the Center of mass energy(c.m.) $\sqrt{s}=800$GeV and integrated 
luminosity L=500fb${}^{-1}$. The scale of non commutativity is assumed to be  $\Lambda = 1$TeV. The limits 
on anomalous couplings of the order $10^{-1}$ from forward backward asymmetries while much stringent limits of the order 
$10^{-2}$  from total cross section are obtained if no signal beyond SM is seen.

\end{abstract}

\maketitle

\section{Introduction}
Triple gauge boson couplings arise in the Standard Model(SM) due to the non Abelian
nature of the theory and thus gives the possibility of exploring the bosonic sector of the
SM. Many of such couplings do not appear in SM at the tree
level and even at higher orders thus they are expected to be very small. Hence the precise measurement of 
these couplings could indicate a signal for new physics beyond the SM even if there is no direct production 
of particles beyond SM spectrum. A linear collider at a center of mass energy
of 800 GeV or more in high luminosity regime provide a unique opportunity to measure such couplings 
with an unprecedented accuracy where we can distinguish such effects from the SM predictions. Moreover the 
availability of initial beam polarization option, can significantly enhance the sensitivity to such 
effects.

In this study we investigated the expected sensitivity of triple gauge boson 
couplings $Z\gamma\gamma$ and $\gamma\gamma\gamma$,  to leading order that will contribute in 
$e^{-}e^{+} \rightarrow \gamma\gamma$ process  at proposed International Linear Collider(ILC)~\cite{ILC1,ILC2}. 
This investigation is done within the frame work of non commutative SM(NCSM).

Quantum field theories constructed on non commutative(NC) space time have been extensively 
explored in the past few years. This field has received much attention due to its possible 
connection with quantum gravity and because of its natural origin in
string theories. Infact Seiberg and Witten\cite{bergwitten} described how
NC gauge theory can emerge as a low energy manifestation of open string theory. This work
has stimulated many paper on non commutative models\cite{snyderUV,Minwalla,Susskind,Alvarez,Jaeckel1,
Jaeckel2}.

Hence keeping in mind the above considerations it is reasonable to investigate field theories,
 and in particular 
the standard model of particle physics on non commutative space time. Here we adopt an
approach based on Seiberg-Witten Map(SWM) popularized by the Munich group\cite{Horvat, Wess1,Wess2,Calmetetal, NGetal, Duplancicetal, ggotheta, Melicetal1,Melicetal2}.

In this approach, to construct the NC extension of the standard model (SM) \cite{Calmetetal,NGetal,Melicetal1,Melicetal2},
 which uses the same gauge group and particle content one expands 
the NC gauge fields in non linear power series of $\Theta$ \cite{bergwitten,Wess1,Wess2}. At face value 
it can be seen from the above map that SW approach leads to a field theory with an 
infinite number of vertices and Feynman graphs thereby leading to an uncontrolled degree of divergence 
inturn giving an impression of complete failure of perturbative renormalization. But over the years a 
number of studies have shown that it is possible to construct anomaly free, renormalizable, and effective 
theories at one loop and first order in $\Theta$ \cite{Bichl1,Martin,Martin-Tam1,Buric1,Buric2,Buric3,Martin-Tam2,Ettefaghi,Buric4}. 
The above mentioned studies provide confidence in using the using NC SW expanded SM for phenomenological 
purposes. However it should be mentioned here that the celebrated IR/UV mixing does not exist in the above $\Theta$ expanded approach. Though this is not a drawback in the scales of our interest there do exist certain phenomena that require all orders of the NC parameter be retained. This led to the so called $\Theta$-exact approach, that is from the exact solutions of the SW equations. The phenomenological consequences of this have been explored in \cite{Horvat2,Horvat3,Horvat4,Horvat5}.

The reason why NC collider phenomenology is interesting, comes from the fact that  the scale of non 
commutativity can be as low as a few TeV\cite{Abbiendipairann,NGetal,BuricZgg,Chaichianetal,Carrolletal,Horvat2,Horvat3,Horvat4,Horvat5}, which is amenable for exploration at  the present or the future colliders.
This has led to a great deal of interest in phenomenology  of the NCSM with SWM. Many phenomenological 
signatures  have been studied by different research groups. These works were mainly
done \cite{Dasetal,Hewett01,Schuppgneutr,Haghighatgneutr,chiaraetal,Najafabadipoltopqur,MahajantbW,IltanZdecays,Deshpandetriplecoup,
Najafabaditopqur,Trampeticpheno,BuricZgg,specketal,Abbiendipairann,Melicquarkonia,Melickpi,Alboteanuothsq,Prasanta1,Prasanta2,wang} with 
unpolarized beams with leading corrections to SM starting from $O(\Theta^2)$. However few studies \cite{ggotheta,sumitNC1,AlboteanuLHC,AlboteanuILC} are 
also done  with corrections at the $O(\Theta)$ in cross section. Some previous studies\cite{NGetal,ggotheta, Deshpandetriplecoup,BuricZgg,AlboteanuLHC,AlboteanuILC} 
have also looked into the sensitivity of anomalous triple gauge couplings at Large Hadron Collider(LHC) and ILC.

In this work we have calculated ($O(\Theta)$) corrections for  pair annihilation  with 
and without initial beam polarization. Here, unlike NCQED case, non commutative effects
at leading order also appear in unpolarized cross section due to the presence of axial vector 
coupling of the Z boson. We have also taken into account the effect of earth's 
rotation\cite{earthroteffs1, earthroteffs2, earthroteffs3, earthroteffs4} on 
observable signals of non commutativity. The effects of Non commutativity is studied on various time averaged 
observables to check the sensitivity of anomalous  couplings. Here we note that this process
has been studied previously\cite{Chensame} with unpolarized initial beams and polarized final states.
Note that observation of final state polarization of photons is not possible
at the high energies of ILC.

We have looked at the sensitivity of the anomalous couplings  at the International Linear
Collider(ILC)\cite{ILC1,ILC2} with realistic beam luminosity of 500 fb${}^{-1}$. The availability 
of longitudinal beam polarization of one or both of the $e^{-}e^{+}$ beams,
can give the opportunity to test  the couplings which otherwise are absent in the observables
with unpolarized initial beam. If these beams become available at future Linear Colliders then it  
will serve as crucial test for these anomalous couplings of NCSM in the process we discuss.

The rest of the paper is organized as follows. In section II, we give the calculational details of our work for the mentioned process. In section III, we will present our numerical results. Finally we conclude with a section on summary of our results.

\section{Cross sections in the Laboratory Frame}
\begin{figure}[ht]
\includegraphics[width=0.60\textwidth]{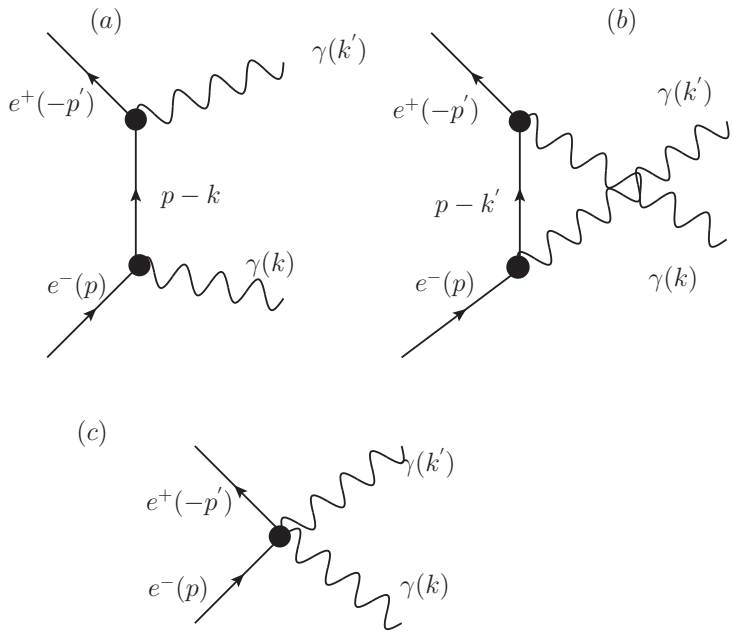}
\caption{{\bf NCQED Diagrams.}
Feynman Diagrams corresponding to NCQED.
\label{NCQEDdiags1}}
\end{figure}

This process in NCQED proceed at the tree level by the following diagrams(Fig.\ref{NCQEDdiags1}).
The first two diagrams also appears in pure QED while 3rd one arises just because of non commutative 
nature of space time and is a contact interaction. 

\begin{figure}[ht]
\includegraphics[width=0.8\textwidth]{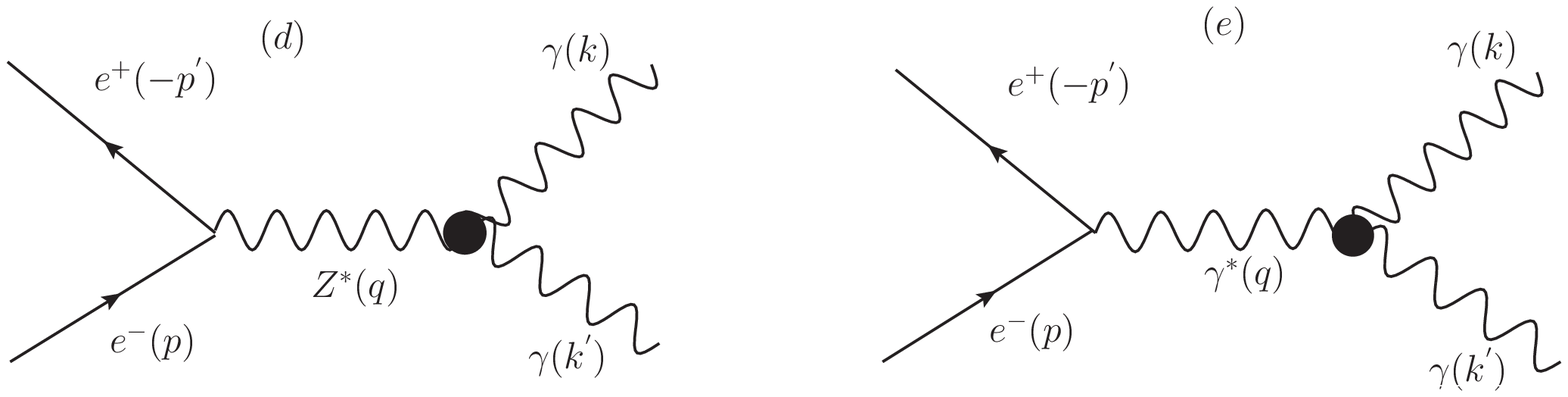}
\caption{{\bf Non Minimal NCSM:}
s-channel Feynman Diagrams with anomalous couplings. Here $q=k+ k^{'}= p+p^{'}$ is the momentum of
the propagator. 
\label{schanlDiagrms}}
\end{figure}

However in non minimal version of Standard model this process also contain two additional 
s-channel diagrams(Fig. \ref{schanlDiagrms}) with anomalous triple gauge boson vertices $Z\gamma\gamma, \gamma\gamma\gamma$.

The squared amplitude for the above process is given by the expression
\[ |A|^2 = |A_{SM}|^2 + (A_{SM})^{*}A_{NC}[O(\Theta)] + A_{SM}(A_{NC})^{*}[O(\Theta)] + |A_{NC}|^2[O(\Theta^2)]\]

Here as mentioned in introduction we are restricting ourselves only to $O(\Theta)$ thus the
interference between SM and NC term can provide required corrections to cross section.

Since non commutative parameter is
considered as elementary constant in nature so its direction is fixed in some non rotating coordinate 
system(can be taken to celestial sphere). However the experiment is done in laboratory coordinate system which 
is rotating with earth's rotation. So one should take into account these rotation effects on $\Theta_{\mu\nu}$ 
in this frame before moving towards the phenomenological investigations. 

\begin{figure}[t]
\begin{center}\lvm
\begin{minipage}[c]{160mm}
\begin{tabular}{ccc}
\begin{minipage}[c]{80mm}
\centerline{
\includegraphics[width=70mm, height=50mm, angle=0]{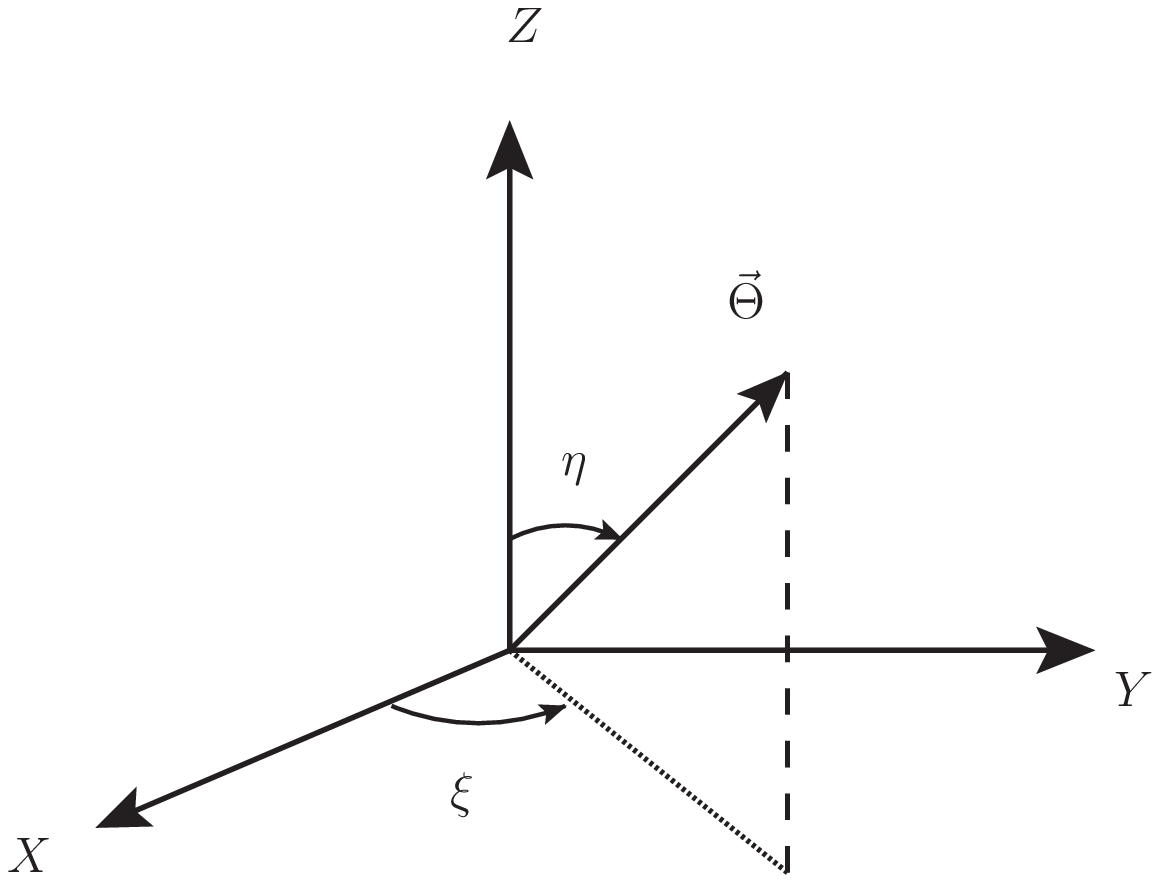}
}\vspace{-0.5cm}
\label{fig:coordinates}
\end{minipage}
&\quad
\begin{minipage}[c]{80mm}
\centerline{
\includegraphics[width=70mm, height=50mm, angle=0]{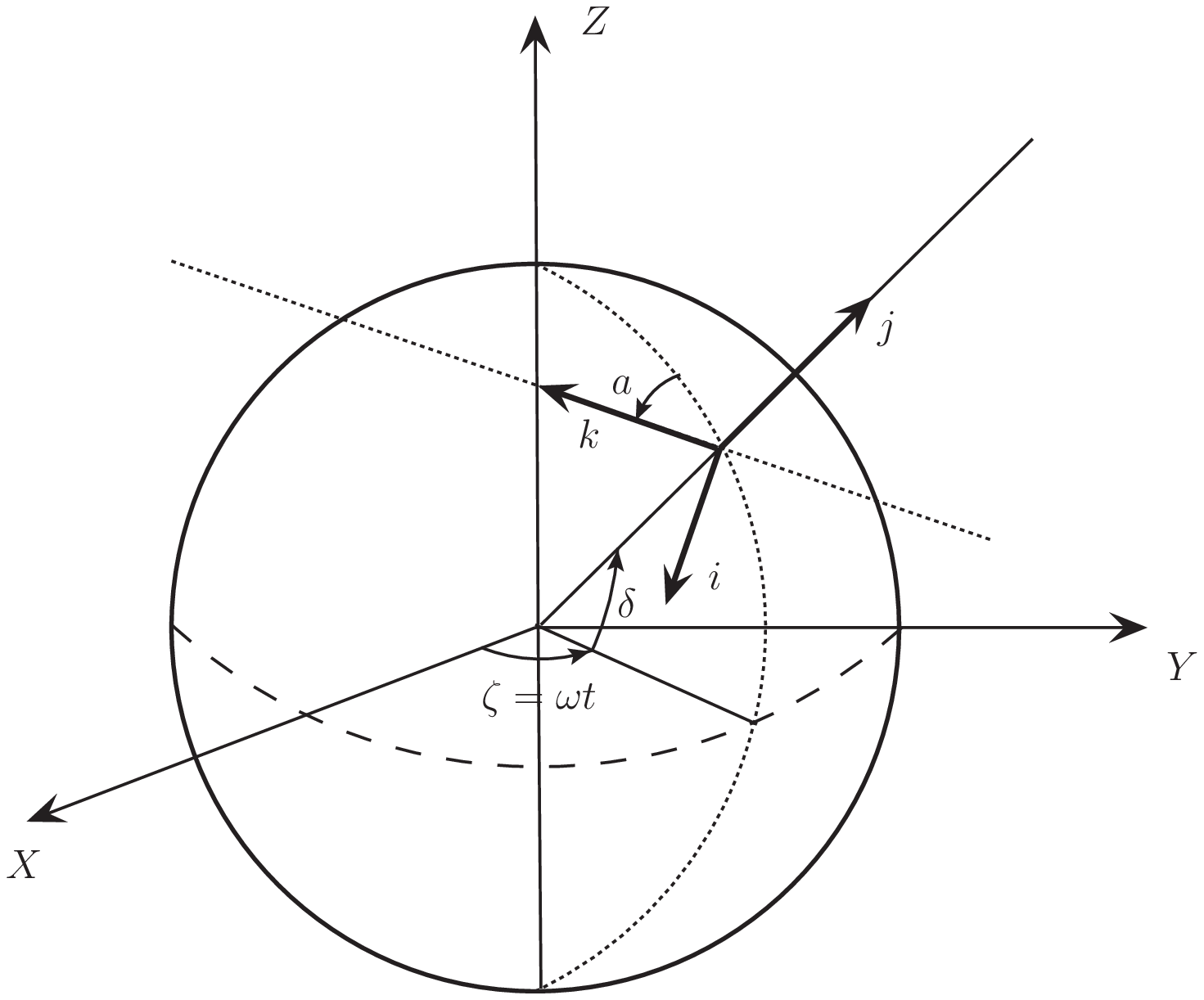}
}\vspace{-0.5cm}
\label{fig:localframe}
\end{minipage}
\end{tabular}
\vspace{0.5cm}
\caption[]{X-Y-Z is the primary coordinate system while $\{ \hat{i}-\hat{j}-\hat{k} \}$ are unit vectors pertaining  
to the laboratory coordinate system. The direction of $\vec{\Theta}$ is defined by angles $\eta$ and $\xi$.}
\end{minipage}
\end{center}
\end{figure}

These effects were considered in many previous studies\cite{earthroteffs1, earthroteffs2, earthroteffs3, earthroteffs4} but 
we are here following the lines of\cite{earthroteffs3}.
In the laboratory coordinate system,
 the orthonormal basis of 
 the non rotating(primary) coordinate system can be written as
\begin{eqnarray}
\begin{array}{ccc}
\overrightarrow{i}= 
   \left(\begin{array}{c}
      c_a s_\zeta + s_\delta s_a c_\zeta \\
      c_\delta c_\zeta \\
      s_a s_\zeta - s_\delta c_a c_\zeta 
   \end{array}\right),
&
\overrightarrow{j}=
   \left(\begin{array}{c}
     -c_a c_\zeta + s_\delta s_a s_\zeta \\
      c_\delta s_\zeta \\
     -s_a c_\zeta - s_\delta c_a s_\zeta 
   \end{array}\right),
&
\overrightarrow{k}=
   \left(\begin{array}{c}
      -c_\delta s_a  \\
       s_\delta  \\
       c_\delta c_a
   \end{array}\right).
\end{array}
\label{eqn:basis}
\end{eqnarray}

Here we have used the abberivations $c_\alpha = \cos \alpha, s_\alpha = \sin \alpha$ etc. ($\delta, a $) defines the location 
of experiment with $-\pi/2 \leq \delta \leq \pi/2$
and $0 \leq a \leq 2\pi$. More details can be found in Ref.~\cite{earthroteffs3}.
 
Thus the NC parameter in the Laboratory frame is given by electric and magnetic components
\bea
\vec{\Theta}_E &=& \Theta_E( \sin \eta_E \cos \xi_E ~\vec{i} +\sin\eta_E \sin\xi_E  ~\vec{j} + \cos\eta_E ~\vec{k} )\nonumber\\
\eea

with
\[ \vec{\Theta}_E = (\Theta^{01},\Theta^{02},\Theta^{03}) \hspace{1cm} \Theta_E = |\vec{\Theta}_E|= 1/\Lambda_E^2 \]

Here  ($\eta, \xi$) specifies the direction of NC parameter($\Theta_{\mu\nu}$) w.r.t primary coordinates system.  
$\Theta_E$ is the absolute values of its electric  components
with corresponding scale $\Lambda_E$.

Our results are based on Feynman rules for NCSM given in Ref.~\cite{Melicetal1,Melicetal2}. For evaluating 
cross section we have used Standard Trace technique and various traces
are obtained by the Mathematica Package FeynCalc\cite{feyncalc}. The trace results 
are also cross checked
in Symbolic Manipulation programme FORM\cite{FORM}. \\

Thus in the Center of Mass frame ($A(p) + B(p^{'}) \rightarrow A(k) + B(k^{'})$)
\bea
p^{\mu} &=& \frac{\sqrt{s}}{2}\{1,0,0, 1\}\nonumber\\
p^{\mu '} &=& \frac{\sqrt{s}}{2}\{1,0,0, -1\}\nonumber\\
k^{\mu } &=& \frac{\sqrt{s}}{2}\{1,\sin\theta\cos\phi, \sin\theta \sin\phi, \cos\theta\}\nonumber \\
k^{\mu '} &=& \frac{\sqrt{s}}{2}\{1,-\sin\theta\cos\phi, -\sin\theta \sin\phi, -\cos\theta\}\nonumber \\
\eea

where $\theta$ is the polar angle and $\phi$ is the azimuthal angle, with initial beam direction
chosen as the z-axis.

Due to the breaking of Lorentz invariance for fixed $\Theta$ background, non commutativity of space time 
lead to dependence of cross section on azimuthal angle which
is absent in Standard Model. The final cross section formulae for different cases are given by:

\subsection{Unpolarized Case}

The differential cross section for $e^{-}e^{+}$unpolarized case
 is given by:

\bea
\displaystyle
\left(\frac{d\sigma}{d\Omega    }\right)_{\overrightarrow{\Theta_E}} &=&
    {\frac{\alpha^2}{ s}}
\left[
      (1 +\cos^2\theta)\csc^2\theta  + \bar{s}_E \{ L_1^{\theta}(\Theta^{02} \cos\phi -  \Theta^{01}\sin\phi) \}
\right]  \; ,
    \label{diffcs1}\nonumber\\
\eea

\bea
L_1^{\theta} &=& 2 C_1 \csc\theta\nonumber\\
C_1 &=& \frac{ C_A}{(1-M_Z^2/s)}K_{Z\gamma\gamma}\nonumber\\\\
\eea

\subsection{Polarized Case}
The differential cross section for $e^{-}$ in Right polarized state is
given by:

\bea
\displaystyle
\left(\frac{d\sigma}{d\Omega    }\right)_{\overrightarrow{\Theta_E}} &=&
    {\frac{\alpha^2}{ s}}
\left[
      (1 +\cos^2\theta)\csc^2\theta  + \bar{s}_E \{ M_1^{\theta}(\Theta^{02} \cos\phi -  \Theta^{01}\sin\phi) \}
\right]  \; ,
    \label{diffcs2}\nonumber\\
\eea

\bea
M_1^{\theta} &=& 2 C_2\csc\theta - \cot\theta\nonumber\\
C_2 &=& \left[ \frac{(C_A - C_V)}{(1-M_Z^2/s)}K_{Z\gamma\gamma}-  K_{\gamma\gamma\gamma}\sin 2\theta_W \right] \nonumber\\\\
\eea

The differential cross section for $e^{-}$ in Left polarized state is
given by:
\bea
\displaystyle
\left(\frac{d\sigma}{d\Omega    }\right)_{\overrightarrow{\Theta_E}} &=&
    {\frac{\alpha^2}{ s}}
\left[
      (1 +\cos^2\theta)\csc^2\theta  + \bar{s}_E \{ N_1^{\theta}(\Theta^{02} \cos\phi -  \Theta^{01}\sin\phi) \}
\right]  \; ,
    \label{diff c.s.}\nonumber\\
\eea

\bea
N_1^{\theta} &=& 2 C_3\csc\theta + \cot\theta\nonumber\\
C_3 &=&  \left[ \frac{(C_A + C_V)}{(1-M_Z^2/s)}K_{Z\gamma\gamma}+ K_{\gamma\gamma\gamma}\sin 2\theta_W \right] \nonumber\\\\
\eea

$C_V$ and $C_A$ are the vector
and axial vector coupling of Z boson with electron and are given by (-1+ 4 $\sin^2\theta_W$)/2 and -1/2 respectively.
Since it is difficult to get time dependent data so one have to take average over full day to be compared with
the experiment. So we will consider here following cross section observables to reveal the effects of non commutativity

\begin{eqnarray}
    \left\langle\frac{d{\sigma}}{d\phi}\right\rangle_{T}
  &\equiv&
  \frac{1}{T_{day}}\int^{T_{day}}_{0}\frac{d\sigma^{}}{d\phi}dt,
\label{eqn:dcr_dphi}
\\
  \langle\sigma\rangle_{T} &\equiv&
  \frac{1}{T_{day}}\int^{T_{day}}_{0}\sigma^{}dt,
\end{eqnarray}
where
\begin{eqnarray}
\frac{d{\sigma}^{}}{d\phi}&\equiv&
  \int^{1}_{-1}\!\!d(\cos\theta)
      \frac{d\sigma^{}}{d\cos\theta{d\phi}},\\
{\sigma}^{}&\equiv&
  \int^{1}_{-1}\!\!d(\cos\theta)
  \int^{2\pi}_{0}\!\!d\phi
      \frac{d\sigma^{}}{d\cos\theta{d\phi}}.
\end{eqnarray}

In addition to these observables, since the terms containing anomalous couplings in cross section flip sign 
under transformation ($\theta \rightarrow (\pi-\theta)$, $\phi \rightarrow (\pi + \phi)$), so we can define 
following forward backward asymmetry(with appropriate cuts on azimuthal angle($\phi$)) which will only be 
sensitive to these couplings.

\begin{eqnarray}
  A_{FB}^{U}(\theta_0)  &=& \frac{\left\langle\sigma_{F}^{f}(\theta_0)\right\rangle_{T}-\left\langle\sigma_{B}^{f}(\theta_0)
\right\rangle_{T}
}
{\sigma_{tot}(\theta_0)} \nonumber\\\\
&=& \frac{  C_1 (4 \theta_0-\pi)\bar{s}_E \sin a \cos\delta \cos\eta}{\pi(\cos\theta_0 +2 \log[\tan\frac{\theta_0}{2}])}\nonumber\\\label{asyeq1}
 \end{eqnarray}

where 
\begin{eqnarray}
 \left\langle \sigma_{F}^{f}(\theta_0)\right\rangle_{T} &=& \frac{1}{T_{day}}\int^{T_{day}}_{0} \left[ \int_{\theta_0}^{\pi/2-\theta_0}\int_{\phi=0}^{\pi}\frac{d\sigma}{d\Omega}\sin\theta d\theta d\phi \right ]dt \nonumber\\\\
 \left\langle \sigma_{B}^{f}(\theta_0)\right\rangle_{T} &=& \frac{1}{T_{day}}\int^{T_{day}}_{0} \left[ \int_{\pi/2 + \theta_0}^{\pi-\theta_0}\int_{\phi=\pi}^{2\pi}\frac{d\sigma}{d\Omega}\sin\theta d\theta d\phi \right ]dt \nonumber\\\\
\sigma_{tot}(\theta_0) &=&  \int_{\theta_0}^{\pi-\theta_0}\int_{\phi=0}^{2\pi}\frac{d\sigma}{d\Omega}\sin\theta d\theta d\phi \nonumber\\
\end{eqnarray}

Similarly one can also define polarized forward backward asymmetry 

\begin{eqnarray}
  A_{FB}^{P}(\theta_0)  &=& \frac{\left\langle\sigma_{FL}^{f}(\theta_0)\right\rangle_{T}-\left\langle\sigma_{BL}^{f}(\theta_0)
\right\rangle_{T}
-(\left\langle\sigma_{FR}^{f}(\theta_0)\right\rangle_{T}-\left\langle\sigma_{BR}^{f}(\theta_0)\right\rangle_{T})}
{\sigma_{tot}(\theta_0)} \nonumber\\\\
&=& \frac{ (C_2-C_3)(\pi-4 \theta_0)\bar{s}_E \sin a \cos\delta \cos\eta}{\pi(\cos\theta_0 +2 \log[\tan\frac{\theta_0}{2}])}\nonumber\\
 \end{eqnarray}

where 
\begin{eqnarray}
 \left\langle \sigma_{FL}^{f}(\theta_0)\right\rangle_{T} &=& \frac{1}{T_{day}}\int^{T_{day}}_{0} \left[ \int_{\theta_0}^{\pi/2-\theta_0}\int_{\phi=0}^{\pi}\frac{d\sigma}{d\Omega}\sin\theta d\theta d\phi \right ]dt \nonumber\\\\
 \left\langle \sigma_{BL}^{f}(\theta_0)\right\rangle_{T} &=& \frac{1}{T_{day}}\int^{T_{day}}_{0} \left[ \int_{\pi/2 + \theta_0}^{\pi-\theta_0}\int_{\phi=\pi}^{2\pi}\frac{d\sigma}{d\Omega}\sin\theta d\theta d\phi \right ]dt \nonumber\\\\
\end{eqnarray}

where L and R refer to the helicity of the incident electron beam; F and B stand for forward
and backward. For unpolarized case, the coefficient of $K_{Z\gamma\gamma}$ in forward backward asymmetry is 
smaller by a factor of $2C_V/C_A$ 
then to its polarized counterpart so the corresponding limit for $K_{Z\gamma\gamma}$ will be
stringent in this case.

\section{Numerical Analysis}
\noindent In this section we will provide the numerical results of our investigation. In order to 
check the sensitivity of anomalous couplings($K_{Z\gamma\gamma}, K_{\gamma\gamma\gamma}$), we studied 
previously defined forward backward asymmetries($A_{FB}$) and total cross section($\left\langle \sigma \right\rangle_{T}$). We fixed the initial beam energy at $\sqrt{s}(=E_{com}) = 800$ GeV. 
The position of Lab system is fixed by taking $\delta = \pi/4$ and $a=\pi/4$. For our sensitivity analysis, 
we assume an integrated luminosity of L=500 fb${}^{-1}$ and we have fixed the NC parameters at $\eta =\pi/4$ and $\Lambda = 1000$ GeV 
while initial phase($\xi$) dependence disappears in time averaged observables.

Here for studying total cross section we applied a cut of ($0-\pi$) on azimuthal angle $\phi$ since non 
commutative effects disappear once we integrate over the full azimuthal
angle($0-2\pi$). Our results are  useful for case $s/\Lambda^2 < 1$ since in this domain one can safely
ignore higher order corrections to cross section.

\begin{figure}[t]
\begin{center}\lvm
\begin{minipage}[c]{160mm}
\begin{tabular}{ccc}
\begin{minipage}[c]{80mm}
\centerline{
\includegraphics[width=80mm, height=85mm, angle=0]{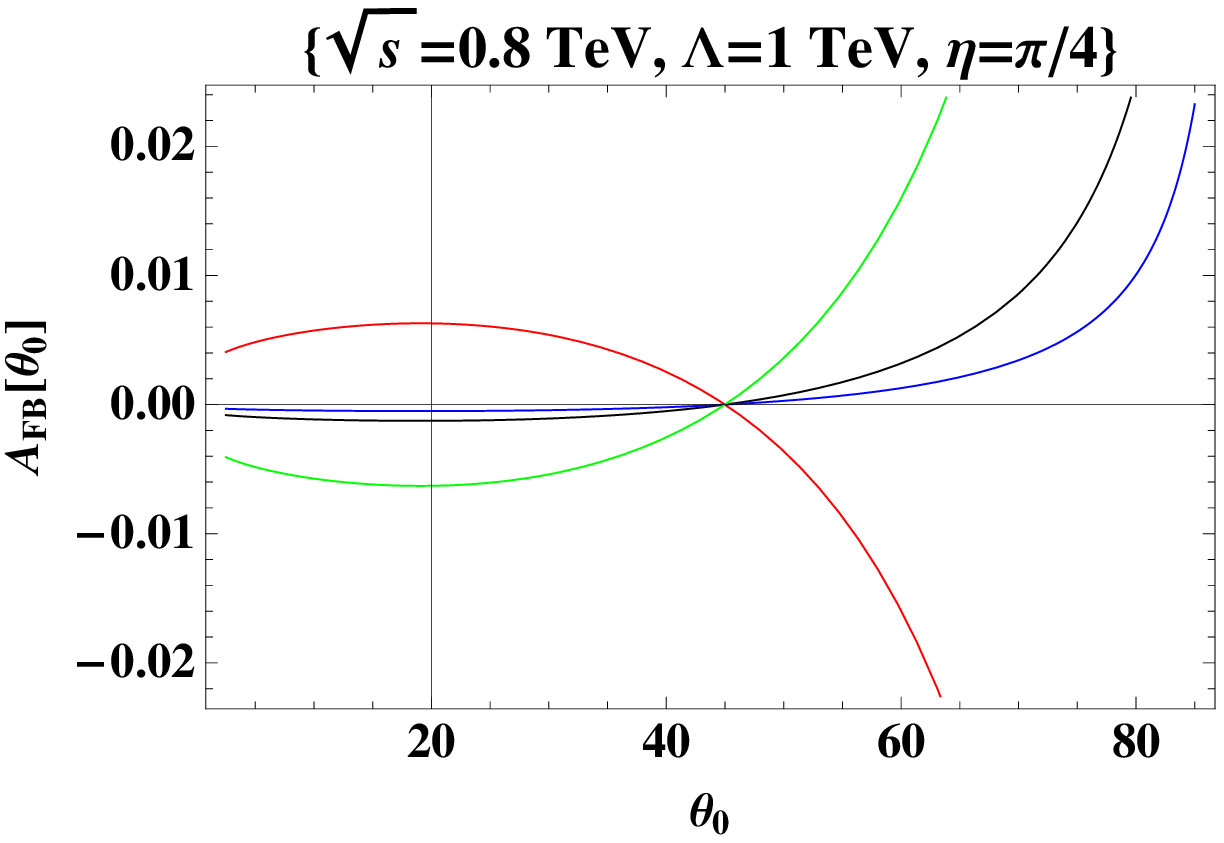}
}\vspace{-0.5cm}
\label{fig:coordinates}
\end{minipage}
&\quad
\begin{minipage}[c]{80mm}
\centerline{
\includegraphics[width=80mm, height=85mm, angle=0]{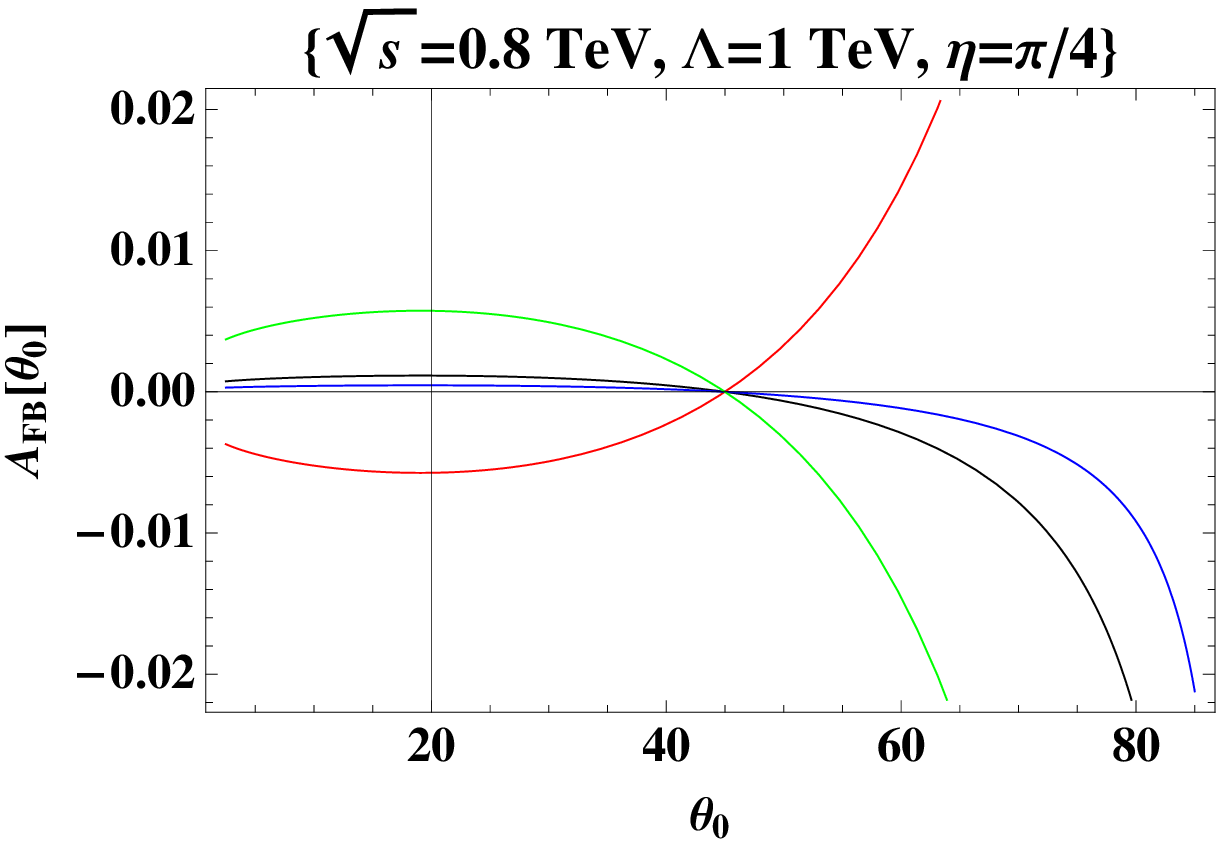}
}\vspace{-0.5cm}
\label{fig:localframe}
\end{minipage}
\end{tabular}
\vspace{-0.9cm}
\caption[]{Forward backward asymmetry for Unpolarized case(Left) and for Polarized
case(Right)vs polar cutoff angle($\theta_0$).
Red, Black, Blue and Green curves corresponds to $K_{Z\gamma\gamma}=$ -0.25, 0.05, 0.02, 0.25 in
unpolarized case and $K_{\gamma\gamma\gamma}=$ -0.25, 0.05, 0.02, 0.25 (with $K_{Z\gamma\gamma}=0$) for polarized case respectively.}\label{Numsect2}
\end{minipage}
\end{center}
\end{figure}

For deriving limits on anomalous couplings we will make use of expressions for  forward backward asymmetry 
along with total cross section. Fig. \ref{Numsect2} shows variation of previously defined forward backward 
asymmetries plotted against polar cutoff angle $\theta_0$ for different values of anomalous couplings. As 
evident from the plots, the magnitude of asymmetries become larger for higher values of polar cutoff 
angle($\theta_0$).  Thus using higher $\theta_0$ will give better limits on couplings. 

The asymmetries are then used to calculate 90\% CL limits with realistic integrated luminosities in
the absence of any signal at ILC. The limit on the coupling at a polar cut off angle($\theta_0$) is related 
to the value of $A(\theta_0)$ of the asymmetry by\cite{limits1,limits2}:
\begin{equation}
 \lambda^{lim}(\theta_0)= \frac{1.64}{|A(\theta_0)|\sqrt{\sigma_{tot}(\theta_0)\cdot L}}\\
\end{equation}

where $|A(\theta_0)|$ is the absolute  value of the asymmetry for unit
value of the coupling. 

\begin{figure}[t]
\begin{center}\lvm
\begin{minipage}[c]{160mm}
\begin{tabular}{ccc}
\begin{minipage}[c]{80mm}
\centerline{
\includegraphics[width=80mm, height=85mm, angle=0]{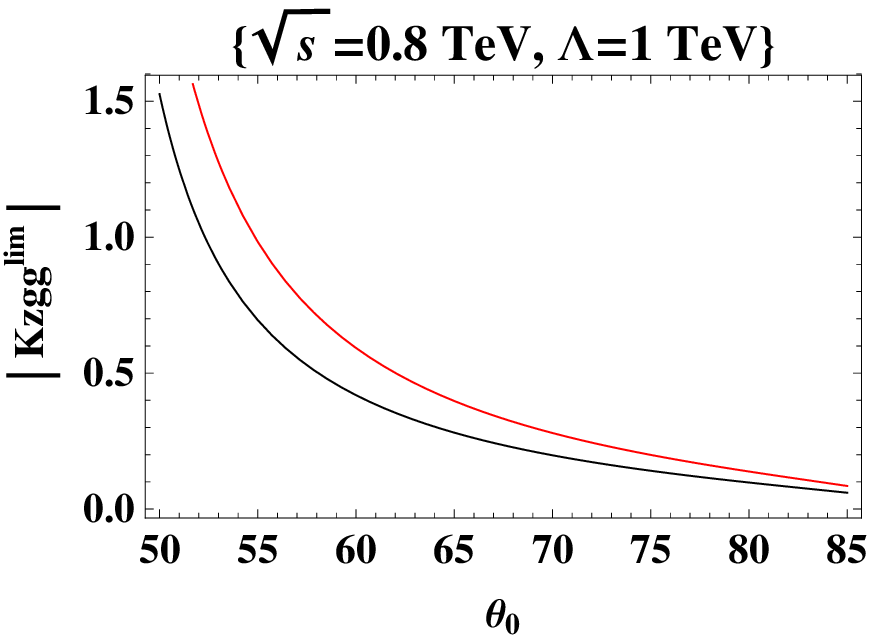}
}\vspace{-0.5cm}
\label{fig:coordinates}
\end{minipage}
&\quad
\begin{minipage}[c]{80mm}
\centerline{
\includegraphics[width=80mm, height=85mm, angle=0]{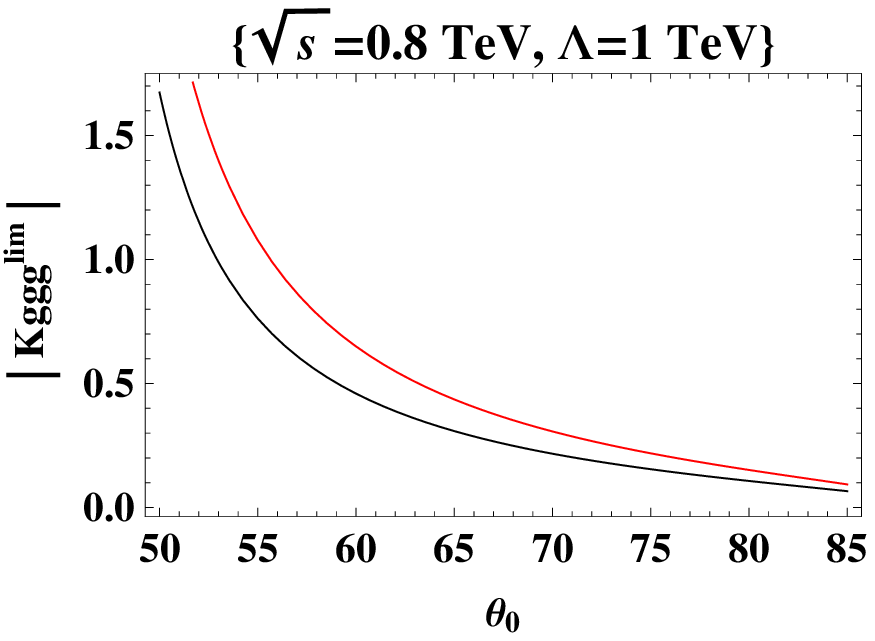}
}\vspace{-0.5cm}
\label{fig:localframe}
\end{minipage}
\end{tabular}
\vspace{-0.9cm}
\caption[]{Limit on $K_{Z\gamma\gamma}$(Left) from unpolarized forward backward asymmetry and on $K_{\gamma\gamma\gamma}$(Right) from polarized forward backward asymmetry. Black and Red curve corresponds to $\eta = \pi/4$ and $\eta = \pi/3$ respectively.}\label{lmAFB}
\end{minipage}
\end{center}
\end{figure}

From Eq. \ref{asyeq1} we see that $A_{FB}^{U}$ solely depends on $K_{Z\gamma\gamma}$, therefore an 
independent limit can be placed on it. However $A_{FB}^{P}$ depends on $K_{Z\gamma\gamma}$
as well as on $K_{\gamma\gamma\gamma}$. Thus for evaluating limit on $K_{\gamma\gamma\gamma}$
we have assumed other anomalous coupling to be zero. From Figs. \ref{lmAFB} it is clear that the best limit 
for $|K_{Z\gamma\gamma}|$ and $|K_{\gamma\gamma\gamma}|$
is achieved for an cutoff angle $\theta_0= 75^{\circ}$.

\begin{figure}[t]
\begin{center}\lvm
\begin{minipage}[c]{160mm}
\begin{tabular}{ccc}
\begin{minipage}[c]{80mm}
\centerline{
\includegraphics[width=80mm, height=85mm, angle=0]{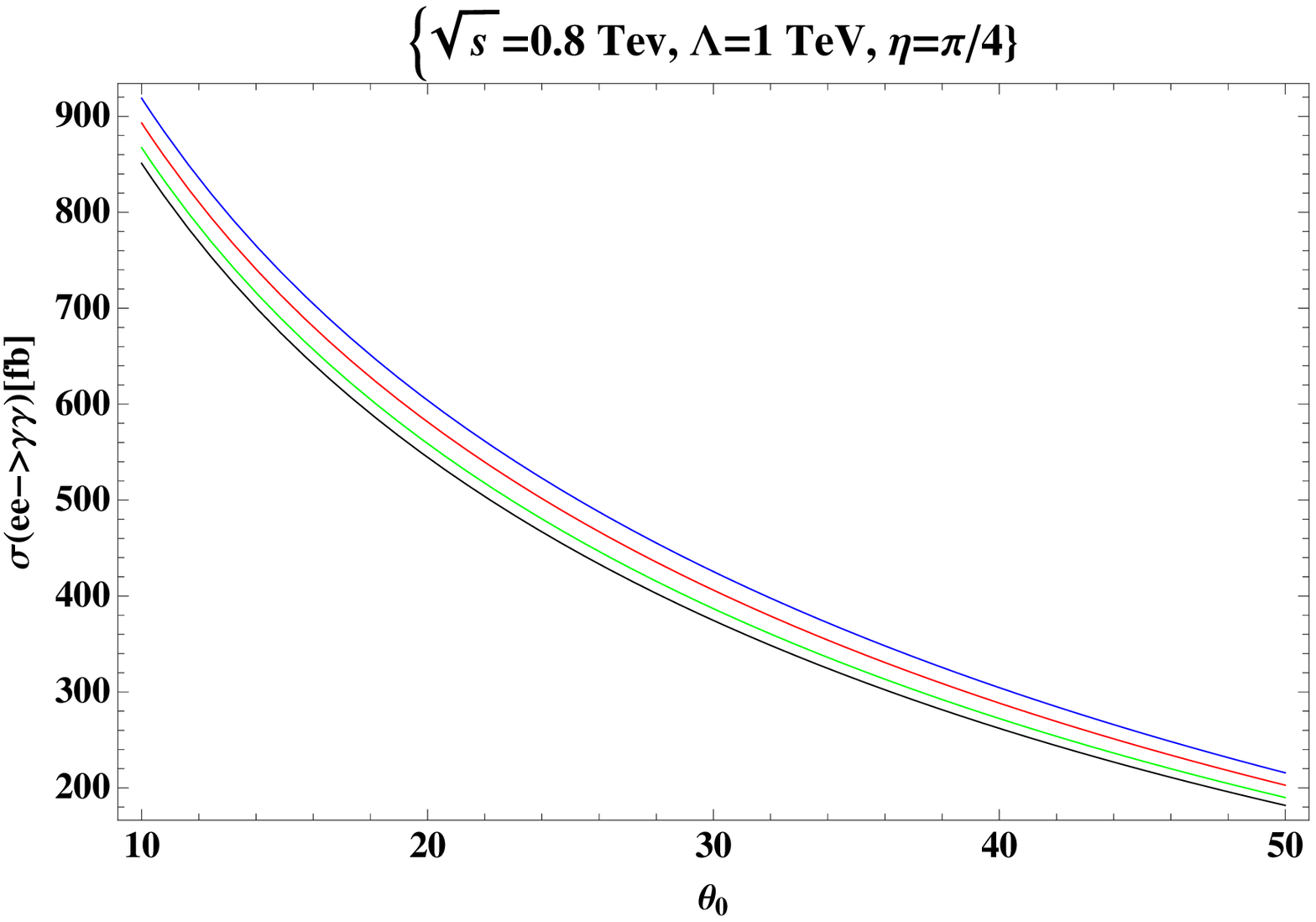}
}\vspace{-0.5cm}
\label{fig:coordinates}
\end{minipage}
&\quad
\begin{minipage}[c]{80mm}
\centerline{
\includegraphics[width=80mm, height=85mm, angle=0]{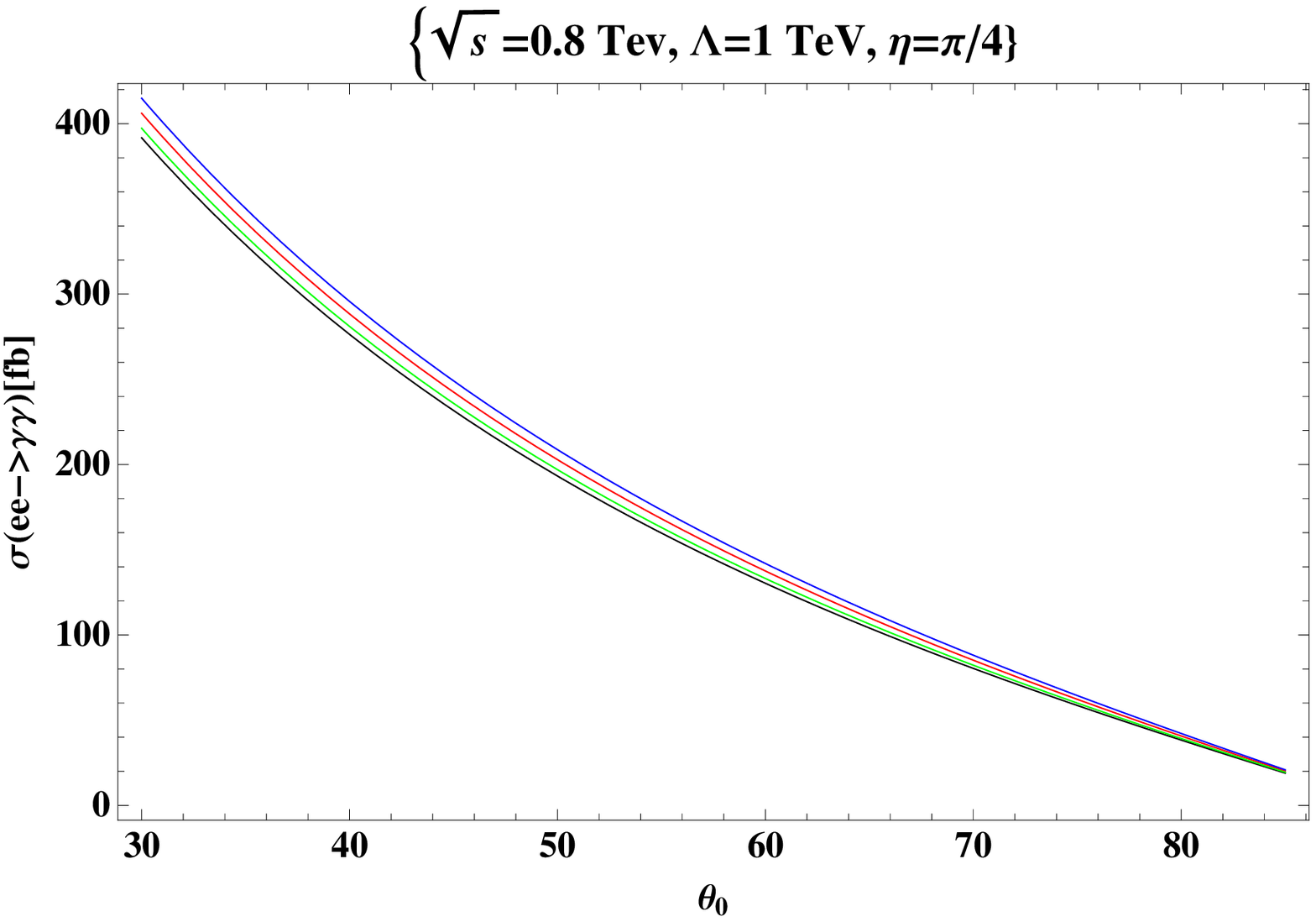}
}\vspace{-0.5cm}
\label{fig:localframe}
\end{minipage}
\end{tabular}
\vspace{-0.9cm}
\caption[]{Time averaged total cross section for Unpolarized 
case with different values of $K_{Zgg}$(Left) and for $K_{\gamma\gamma\gamma}$(Right). Red, Black, Green and Blue curves
corresponds to $K_{Z\gamma\gamma}=$ 0.0, 0.9, 0.55, -0.55 in
unpolarized case and $K_{\gamma\gamma\gamma}=$ 0.0, 0.9, 0.55, -0.55 (with $K_{Z\gamma\gamma}=0$) for polarized case respectively.}\label{Numsect3}
\end{minipage}
\end{center}
\end{figure}

Following the same procedure one can obtain limits from total time averaged cross section. In 
Fig.~\ref{Numsect3} we have plotted the total time
averaged cross section section for different values of anomalous couplings. Here unlike
asymmetries significant deviation from SM case is obtained  for smaller
values of cut-off angle($\theta_0$). We have also
derived the limits on these couplings in case of no excess in signal events is observed at ILC. 

However here the limits are obtained by using the condition that the excess number of events
beyond expected from SM should be smaller than the statistical error in the number of SM events.
This translates to $ L|\sigma_{NP}(\theta_0)|  <  1.64 \sqrt{\sigma_{tot}(\theta_0)\cdot L}$ where
$\sigma_{NP}$ is the NC contribution to the total cross section and L is
the integrated luminosity which we assumed to be 500 fb${}^{-1}$ for current study.

\begin{figure}[t]
\begin{center}\lvm
\begin{minipage}[c]{160mm}
\begin{tabular}{ccc}
\begin{minipage}[c]{80mm}
\centerline{
\includegraphics[width=80mm, height=85mm, angle=0]{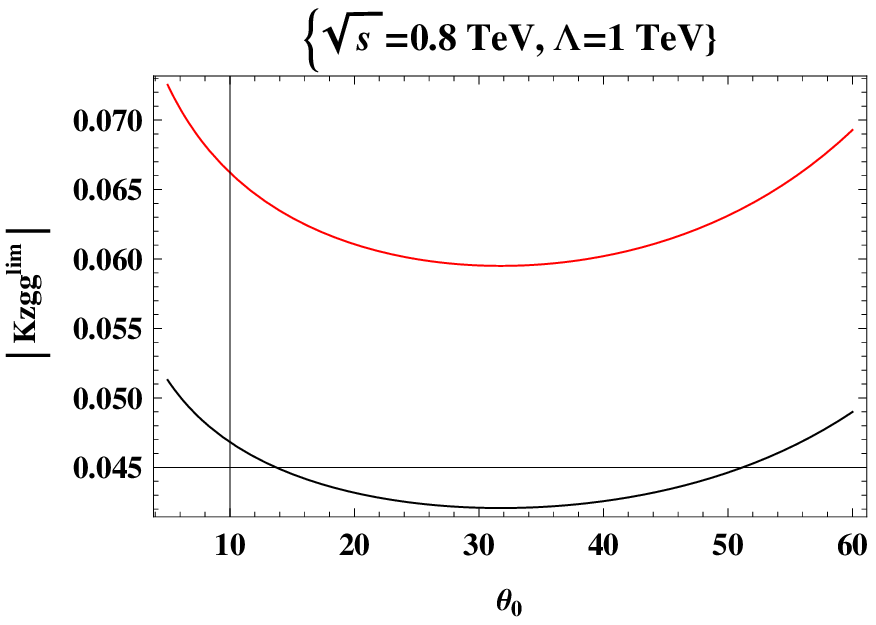}
}\vspace{-0.5cm}
\label{fig:coordinates}
\end{minipage}
&\quad
\begin{minipage}[c]{80mm}
\centerline{
\includegraphics[width=80mm, height=85mm, angle=0]{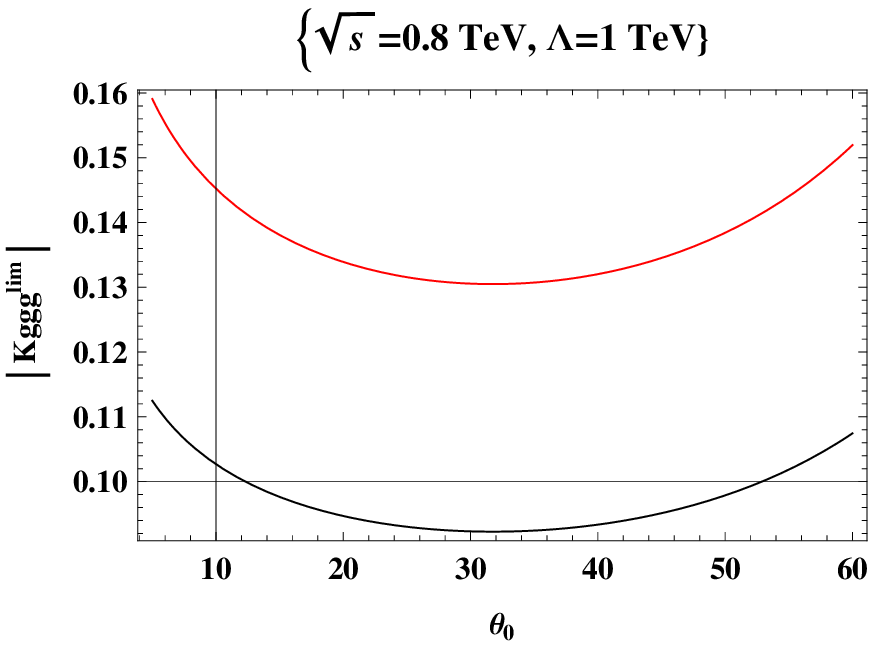}
}\vspace{-0.5cm}
\label{fig:localframe}
\end{minipage}
\end{tabular}
\vspace{-0.9cm}
\caption[]{Limit on $K_{Z\gamma\gamma}$(Left) from unpolarized total cross section and on $K_{\gamma\gamma\gamma}$(Right) from polarized cross section. Black and Red curve corresponds
to $\eta = \pi/4$ and $\eta = \pi/3$ respectively.}\label{LmKtotcrss}
\end{minipage}
\end{center}
\end{figure}

From Eqs.\ref{diffcs1}, \ref{diffcs2} we see that $\sigma_T$ for unpolarized case solely depends on 
$K_{Z\gamma\gamma}$, therefore an 
independent limit can be placed on it. However $\sigma_T$ for unpolarized case depends on $K_{Z\gamma\gamma}$
as well as on $K_{\gamma\gamma\gamma}$. Thus for evaluating limit on $K_{\gamma\gamma\gamma}$
we have assumed other anomalous coupling to be zero. From Figure \ref{LmKtotcrss} it is clear that the 
best limit for $|K_{Z\gamma\gamma}|$ and $|K_{\gamma\gamma\gamma}|$
is obtained for an cutoff angle $\theta_0= 30^{\circ}$.

\begin{table}[!htb]
\centering
\begin{tabular}{|l|c|c|c|c|}
\hline
Coupling & \multicolumn{2}{|c|} {Limits-Unpolarized case}
 & \multicolumn{2}{|c|} {Limits-Polarized case} \\ \cline{2-5}
& $\left\langle \sigma \right\rangle_{T}$ & $A_{FB}$ & $\left\langle \sigma \right\rangle_{T}$ & $A_{FB}$ \\ 
\hline 
$|K_{Z\gamma\gamma}|$ & $4.2\times 10^{-2}$ &$1.4\times 10^{-1}$ & $4.5\times 10^{-2}$ &$9.2\times 10^{-1}$ \\
$|K_{\gamma\gamma\gamma}|$ & - &- & $9.2\times 10^{-2}$ &$1.5\times 10^{-1}$ \\
\hline
\end{tabular}
\caption{90 \% CL limits on the couplings from $\left\langle \sigma \right\rangle_{T}$ for a cut-off
angle of 32$^{\circ}$ and $A_{FB}$ for a cut-off
angle of 75$^{\circ}$. These limits are derived for  $\sqrt{s} = 800$ GeV, $\Lambda = 1000$ GeV and integrated luminosity, L= 500 ${\rm fb}^{-1}$.}
\label{table1}
\end{table} 

In Table \ref{table1} we have quoted the derived limits from asymmetries as well as from time
averaged total cross section. The asymmetries give a limit of order $10^{-1}$ while from total cross
section limits are much more stringent of the order of $10^{-2}$. Thus total cross section is proven
to be much useful observable then forward backward asymmetries.


\section{Summary }
The extension of SM to NC space time with motivations 
coming from string theory and quantum gravity provides interesting phenomenological implications 
since scale of non commutativity could be as low 
as a few TeV, which can be explored at the present or future colliders. In the present work we focused
on exploring the sensitivity of anomalous couplings($K_{Z\gamma\gamma}, K_{\gamma\gamma\gamma}$) that
will contribute to the process $e^{+}e^{-}\rightarrow \gamma\gamma$ process at ILC. 

We have done our study with unpolarized as well as taking into account the initial 
beam polarization effects.  We restricted ourselves to the leading order effects of
non commutativity to be occurred at leading order in $\Theta$(i.e. O$(\Theta)$) at cross section
level. Unlike NCQED case non commutative effects at O$(\Theta)$
also appear in unpolarized cross section due to the presence of axial vector coupling of Z boson.  

In this 
analysis we have also taken into account the apparent time variation
of non commutative parameter($\Theta^{\mu\nu}$) in Laboratory frame. We have used time averaged
observables for this study.

The NC corrections to the considered process are sensitive to the 
electric component($\vec{\Theta}_E$) of NC parameter ($\vec{\Theta}$). However for checking the sensitivity
of anomalous couplings at ILC we  used time averaged total
cross section and forward backward asymmetries as observables. This analysis is done under
realistic ILC conditions with the Center of mass energy(c.m.) $\sqrt{s}=800$GeV and integrated luminosity 
L=500fb${}^{-1}$. The scale of non commutativity is assumed to be  $\Lambda = 1$TeV and Lab coordinates are 
fixed to be $(\delta, a)=(\pi/4, \pi/4)$. 

The observables for unpolarized case are only sensitive to $K_{Z\gamma\gamma}$ while for polarized
case they are sensitive to both couplings. For putting limits from polarized case we have assumed
one coupling to be zero at a time. The asymmetries give a limit of order $10^{-1}$ while limits from 
total cross section are much more stringent, of the order of $10^{-2}$ on absolute value of the anomalous
couplings. 
 

\begin{acknowledgments}
S.K.G is grateful to Dr. Rolf Mertig for useful correspondence. The work of N.G.D is supported by
the US DOE under Grant No. DE-FG02-96ER40969.
  S.K.G acknowledges partial support from DST Ramanujan Fellowship
SR/S2/RJN-25/2008.\\

\end{acknowledgments}



\end{document}